# Le groupe de Lorentz-Einstein

# comme conséquence d'un quantum d'action

The Lorentz-Einstein group as a consequence of a quantum of action

A. Avramesco, le Château, 25290 Ornans

<a.avramesco@wanadoo.fr>

*Résumé :* on montre comment la quantification naturelle des intervalles entre évènements (au sens relativiste) conduit à la définition de deux unités, *c* et *h*, entraînant aussi bien les formules de transformation de coordonnées que les formules quantiques fondamentales.

*Abstract :* it is shown that the natural quantization of intervals between events (in the relativistic sense) leads to the definition of two units, *c* and *h*, implying the formulae for the transformation of co-ordinates as well as the fundamental quantum formulae.

*Lors du passage de la lecture initiale de ce texte, en **Word PC 6**, à la lecture en **Word Mac 2001**, sont apparues certaines distorsions (la plus grave a été le remplacement, dans toutes les équations du §4, p.9, des coefficients 1/2 par un tiret). L'auteur a corrigé tout ce qu'il a aperçu de ces distorsions : il présente ses excuses aux lecteurs pour ce qui en demeurerait au moment de la mise en ligne.*

Lecteur pressé, adieu.

*[...] les opinions des hommes sont receues à la suitte des creances anciennes, par authorité et à credit, comme si c'estoit religion et loy. On reçoit comme un jargon ce qui en est communement tenu ; on reçoit cette verité avec tout son bastiment et attelage d'argumens et de preuves, comme un corps ferme et solide qu'on n'esbranle plus, qu'on ne juge plus. Au contraire, chacun, à qui mieux mieux, va plastrant et confortant cette creance receue, de tout ce que peut sa raison, qui est un util souple, contournable et accommodable à toute figure. Ainsi se remplit le monde et se confit en fadesse et en mensonge.*

(Montaigne, *Essais*, livre II, chap. XII, *Apologie de Raimond Sebond* ; cette *Apologie* emplit, dans l'édition *Classiques Garnier* de 1952, les pages 115 à 317, et la citation ci-dessus se lit p. 238 : comme il est souvent difficile de trouver une édition particulière de textes fondamentaux, comme celui-là, et qu'il est très utile d'inviter le lecteur à feuilleter et méditer autour de quelques phrases choisies, les références ci-dessous seront données dans le même esprit, c'est-à-dire qu'il faut un certain effort pour y accéder, et donc regarder un peu autour. Car il s'agit d'inviter à la réflexion scientifique, sur des sujets qui exigent une culture, une connaissance historique et une longue interrogation sur les fondements des formalismes et théories acquis. Cela ne peut se faire que si on reprend et remédite longuement par soi-même des textes absolument essentiels. On propose donc tout le contraire d'un carnet mondain faussement précis, mais soigneusement mis à jour, de relations dans des comités de revues...)

## Références

[1] « La question de la validité des hypothèses de la géométrie dans l'infiniment petit est liée à la question du principe intime des rapports métriques dans l'espace. Dans cette dernière [...], on trouve l'application de la remarque [...] que, dans une variété discrète, le principe des rapports métriques est déjà contenu dans le concept de cette variété, tandis que dans une variété continue ce principe doit venir d'ailleurs. Il faut donc, ou que la réalité sur laquelle est fondé l'espace forme une variété discrète, ou que le fondement des rapports métriques soit cherché en dehors de lui, dans les forces de liaison qui agissent en lui. La réponse à ces questions ne peut s'obtenir qu'en partant de la conception des phénomènes vérifiée jusqu'ici par l'expérience [...], et en apportant à cette conception les modifications successives exigées par les faits qu'elle ne peut expliquer. Des recherches partant de concepts généraux [...] ne peuvent avoir d'autre utilité que d'empêcher ce travail d'être entravé par des vues trop étroites, et que le progrès dans la connaissance de la mutuelle dépendance des choses ne trouve un obstacle dans les préjugés traditionnels. » (Riemann, *Sur les hypothèses qui se trouvent à la base de la géométrie,* thèse d'habilitation de 1854, à retrouver par exemple dans la trad. fr. chez Gauthier-Villars 1898, à partir de p. 280 et spécialement p. 297).

[2] « Il est digne de noter que l'énergie et la fréquence d'un complexe de lumière se modifient d'après les mêmes lois avec l'état de mouvement de l'observateur. » (Einstein, *Sur l'électrodynamique des corps en mouvement*, texte de 1905 repris chez Gauthier-Villars dans la traduction de Maurice Solovine, Paris, 1965, p. 37, partie *Electrodynamique*, première moitié du paragraphe 8).

[3] « On ne remarqua pas que le véritable élément de la <u>description</u> spatio-temporelle est l'évènement, qui est <u>décrit</u> dans l'espace et le temps par les quatre nombres $x_1$, $x_2$, $x_3$, t. » (Einstein, *Quatre conférences sur la théorie de la relativité* de 1921, éd.fr. Gauthier-Villars 1955, p. 27, paragraphe sur le *Principe de constance de vitesse de la lumière*, avant l'explicitation de la *transformation de Lorentz* ; les <u>soulignements</u> ne sont pas dans l'original).

[4] « La mesure d'une partie quelconque d'une multiplicité discrète est donnée par le nombre des éléments qu'elle contient. Ainsi une multiplicité discrète porte le principe de sa métrique, en soi a priori, comme Riemann le dit. » (Hermann Weyl, *Temps, espace, matière*, éd. fr. chez Albert Blanchard 1922, p. 83 ; mais surtout, quoique citant aussi longuement le texte [1] de Riemann, Weyl souligne aussitôt qu'avec les quanta, la <u>première</u> hypothèse de Riemann, celle relative à une <u>variété discrète</u> comme <u>fondement</u> d'un espace, prend des accents singulièrement profonds).

[5] [trad.] « Je considère comme improbable que les relations <u>individuelles</u> d'intervalles entre points-évènements suivent une telle règle définie [une métrique déterminée]. Un examen microscopique montrerait probablement qu'elles sont tout à fait arbitraires, les relations entre points soi-disant intermédiaires n'étant pas nécessairement intermédiaires. Peut-être même l'intervalle primitif n'est-il pas quantitatif, mais simplement 1 pour certains couples de points-évènements et 0 pour d'autres [...] De là le point de départ de la variété sans fin de la nature. » (Eddington, *Space, time and gravitation*, édition de Cambridge 1920, p. 188).

[6] « Ce n'est pas l'espace et le temps, concepts statistiques, qui peuvent nous permettre de décrire les propriétés des entités élémentaires [...] La véritable physique quantique serait sans doute une physique qui, renonçant aux idées de position, d'instant, d'objet et à tout ce qui constitue notre intuition usuelle, partirait de notions et d'hypothèses purement quantiques [...] » (L. de Broglie, *Continu et discontinu en physique moderne*, éd. Albin Michel, Paris, 1943, p. 72).

[7] **« On peut donner de bonnes raisons pour lesquelles la réalité ne peut pas être du tout représentée par un champ continu. Des phénomènes quantiques paraît résulter avec certitude qu'un système fini, d'énergie finie, peut être complètement décrit par un ensemble fini de nombres (nombres quantiques). Cela ne semble pas être en conformité avec une théorie continue, et doit mener à un essai de découverte d'une théorie purement algébrique de la réalité. Mais personne ne sait comment obtenir la base d'une telle théorie. »** (Einstein, appendice de décembre 1954 aux *Quatre conférences* citées plus haut, dans la traduction de Marie-Antoinette Tonnelat et Maurice Solovine tirée à part en 1955, paragraphe final D).

*L'étude ci-dessous montre comment une <u>mathématique purement quantique</u>, celle de graphes orientés acycliques, répond aux prémonitions de la lignée dont on vient de repérer quelques jalons marquants, et notamment aux indications du testament scientifique d'Einstein relu à l'instant. On retrouvera ainsi comme <u>résultats</u> aussi bien les transformations "de Lorentz" (pourtant jusqu'ici considérées comme type de transformations continues) que les formules quantiques de base sur énergie et impulsion, au lieu d'articuler plus difficilement données expérimentales et mise en forme théorique. L'ennemi principal dans la rédaction comme dans la compréhension de cette étude est la référence aux « creances anciennes » comme dit Montaigne, aux « préjugés traditionnels » comme dit Riemann, spécialement à la manie qu'entre deux valeurs de grandeurs essentielles en physique on puisse toujours donner un sens à une valeur intermédiaire : autrement dit la manie continuiste, devenue dogme en « natura non facit saltus ». Même les réflexes spatio-temporels sont moins graves.*

*Pour ce qui concerne la forme adoptée ici, après bien des exposés et rédactions, elle cherche surtout à éviter la reprise des schémas classiques : car le glissement l'une contre l'autre de deux rangées d'atomes équirépartis ne présente pas de difficulté particulière dans ces schémas, et cette référence suffit si on se donne le mal de la reconstituer — plutôt que de la retrouver, partiellement et dans un esprit tout différent, dans quelque recueil d'exercices —. On voit alors comment la nouvelle description se passe des conceptions habituelles. Ceci posé, le lecteur demeure libre de refuser l'effort pour lire et comprendre ; mais il faut rappeler que, dans le cadre de la science, ce refus suffit à interdire le jugement : lecteur honnête, pardon de ce rappel.*

# 1. Introduction

Dès les débuts des expositions relativistes donc, la profondeur des concepts d'évènements et d'intervalles s'avérait supérieure à celle des descriptions, des *situations* en termes de coordonnées, tandis que s'éclairait la relation entre intervalle et action : ainsi dès Langevin, on remarquait que la mécanique relativiste résultait d'un principe de moindre action indépendant de l'électrodynamique (cf. par exemple l'utilisation de la proportionnalité entre intervalle et action exposée par Landau et Lifchitz dans leur cours de Physique théorique, tome II, théorie du champ, chapitre II). En un sens, ce qui suit n'est donc qu'une prise en compte de l'expérience, mais à un niveau essentiel : celui du courant principal des condensés d'expériences, que constitue l'histoire, la succession des mises en forme physiques.

Les résultats ci-après sont restreints à deux dimensions, qui s'interprètent l'une en espace et l'autre en celle liée au temps. Cette restriction n'est nullement dans la nature des idées proposées : elle simplifie seulement une première exposition.

Soient alors les données de la figure 0. En lecture spatio-temporelle, celle-ci schématise les évènements successifs correspondant à une droite physique constituée de points matériels régulièrement répartis : chaque atome ou "point matériel" échange avec ses deux voisins des signaux immédiatement réfléchis — les émissions-absorptions sont instantanées, et simultanées pour tous les points d'abscisse paire (donc, à instants décalés, simultanées aussi pour tous les points d'abscisse impaire) : ainsi P étant pris pour origine, la ligne $t = 0$ est la "verticale" passant par P, et la trajectoire du point correspondant est l'"horizontale" d'abscisse (réduite) $x/c = 0$. Il est clair que prétendre représenter une droite physique par d'aussi simples échanges de signaux entre atomes ponctuels est grossièrement simplificateur : il n'en est, comme on verra, que plus remarquable que cela suffise à retrouver les règles de transformations de coordonnées.

Car cette même figure 0 peut être lue suivant une tout autre grille. On peut la voir comme un graphe orienté acyclique où un sommet est un évènement, et où deux évènements constituent ou (exclusivement) ne constituent pas un couple orienté dit flèche. Il est capital de saisir que chaque couple orienté, chaque flèche est constituée de deux évènements ordonnés, et non pas d'un arc dans un espace de plongement : chacune des flèches est un quantum. Tout de suite après ce fondement, le fait que les couples soient orientés est lui aussi fondamental : l'irréversibilité est donc au fond des choses — au contraire des habitudes acquises.

Ceci posé, la topologie du graphe fait distinguer deux sortes de flèches : l'une appelée ici, d'après les figures, ascendante, l'autre descendante. On repère alors un sommet quelconque à partir d'un sommet P pris pour origine de la façon suivante : P étant affecté de (0, 0), une flèche ascendante A augmente d'une unité les deux entiers : (+1, +1), une flèche descendante D augmente le premier mais abaisse le second, également d'une unité : (+1, -1). Ainsi, le premier entier augmente toujours en suivant le cours des flèches, et doit donc être associé au temps du graphe-repère, $t$ ; le second entier augmente le long d'un des types de flèches, et diminue le long de l'autre : c'est l'abscisse (réduite) $x/c$, le rapport de réduction étant égal à 1 dans les définitions précédentes. On notera ce rapport $c$ pour marquer qu'il correspond de façon essentielle à deux décomptes distincts, on remarquera bien que le couple d'entiers (q, r) est ainsi déterminé, comme indiqué sur la figure 0, et on enregistrera quatre sommets du type (q,r), (q+1, r), (q, r+1) et (q+1, r+1) comme *pavé élémentaire*.

Dans cette vue, le graphe avec sa topologie est structure fondamentale, les représentations pouvant être déformées de manière quelconque (comme si elles étaient dessinées sur caoutchouc, pour reprendre l'image la plus parlante) sans rien y changer :
voilà le **pivot**.

Il faut s'attarder sur ce basculement. A partir du graphe, notion essentielle où figurent seulement, du point de vue physique, les évènements et leurs relations **quantiques** — du point de vue mathématique, les sommets et leurs relations **quantiques** —, la notion d'espace est un résultat. L'espace est fondé sur une variété discrète, comme le dit Riemann, ce n'est pas l'espace qui est discrétisé : ce sera plus net avec les comparaisons de repères ou systèmes physiques, comme on le verra au paragraphe suivant. En outre, il est déjà naturel d'arrêter en quelque sorte le graphe au moins à un ensemble de sommets "pendants", c'est-à-dire de considérer que l'on sait construire indéfiniment d'autres sommets du graphe à partir d'un ensemble donné si on connaît son programme (pour le moment tout simple : autour de chaque sommet arrivent, et partent, deux flèches, une ascendante et une descendante) ; mais il n'y a aucune raison d'arrêter l'ensemble donné à une ligne $t = constante$, ni de considérer que les sommets postérieurs (en un sens évident) sont autres que d'abord virtuels, à réaliser, à partir de ceux qui sont déjà donnés. La distinction entre passé et futur est déjà un peu plus radicale que dans les géométries d'univers habituelles. Ceci est essentiel en vue d'autres extensions.

On conservera cependant ici des géométries où les flèches sont inclinées d'angle $\pm \pi/4$ en vue de la traduction spatio-temporelle : car si les lignes coordonnées ne sont en général pas orthogonales, les bissectrices $x/c = \pm t$, elles, le sont.

## 2. Transformations de Lorentz-Einstein

La figure 0 schématisant donc une droite physique, quel doit être le schéma pour le glissement l'une contre l'autre de deux telles droites ?

On peut ici soit revenir à la description classique et parler d'abord de coïncidences entre atomes équirépartis sur chaque droite, soit accepter l'effort de transcrire immédiatement dans les termes nouveaux : l'exposé ci-dessous s'en tient largement à ces termes, comme indiqué en introduction.

Puisque chaque sommet d'un graphe-repère correspond à un évènement, on doit faire correspondre, à un évènement commun à deux repères, un sommet commun aux deux graphes : on appellera cela une coïncidence. Si alors le glissement d'une droite contre l'autre est uniforme, et si on respecte la symétrie avant et arrière pour chacune, chaque coïncidence doit pouvoir être prise pour origine sans modifier le schéma d'ensemble. La double périodicité du tableau (fig. 1) des coïncidences en résulte.
(N. B. : les entiers (n, p) indiquant les *coïncidences*, suivant la figure, ne sont pas, bien sûr, les entiers (t, x/c) numérotant des *sommets* quelconques, car il y a beaucoup plus de *sommets* ordinaires que de *coïncidences* : le lien simple entre les deux doublets va justement être examiné).

Un des graphes-repères (appelé par exemple : rouge) peut être représenté à partir de la *maille élémentaire* de la figure 2. Car d'abord, la répétition (doublement indéfinie) de cette maille donne bien un graphe identique à celui de la fig. 0 : la représentation géométrique seule change, la structure, la topologie, n'a pas varié, comme on y a insisté en présentant le **pivot** (pour revenir aux termes classiques : chaque "atome" de la droite rouge échange avec ses voisins les mêmes "signaux" dans les mêmes conditions, seul le repère est différent). Ensuite, la périodicité des coïncidences est de 1 en *n* et *p* : plus précisément mais lourdement, le rôle de $C_{n+1, p}$ vis-à-vis de $C_{n, p}$ est semblable à celui de $C_{n+1, p+1}$ vis-à-vis de $C_{n, p+1}$, et de même le rôle de $C_{n, p+1}$ vis-à-vis de $C_{n, p}$ semblable à celui de $C_{n+1, p+1}$ vis-à-vis de $C_{n+1, p}$ ; donc la *maille élémentaire*, entre coïncidences, du graphe rouge définit deux entiers $N_<$ et $N_>$, correspondant aux nombres de flèches rouges sur chaque côté de la maille. Alors, la réciprocité relativiste exigeant la symétrie des rôles, le second repère ("bleu") aura, *entre les mêmes coïncidences*, la *maille élémentaire* de la figure 3 par simple permutation de $N_<$ et $N_>$ (classiquement : "si, sur la droite rouge, et dans le sens de son mouvement, un signal parcourt $N_<$ intervalles entre atomes pour $N_>$ sur la bleue, il doit en être de même pour la bleue par rapport à la rouge" ; mais le langage des graphes, intervalles au sens relativiste et évènements, est comme on voit plus simple, plus physique, et donc préférable).

De là, il est immédiat d'évaluer les valeurs (t, x/c) attachées à un sommet-évènement, telles qu'elles ont été définies au titre **1**. En affectant l'indice $_r$ au graphe rouge, $_b$ au graphe bleu, on a pour $C_{n, p}$ :

$$t_r = n N_< + p N_>, \quad x_r / c = n N_< - p N_>,$$
$$t_b = n N_> + p N_<, \quad x_b / c = n N_> - p N_<.$$

La vitesse (réduite) $w = v/c$ du graphe repère rouge par rapport au bleu est le rapport d'accroissements $(\Delta x_b /c) / (\Delta t_b)$ à abscisse rouge fixée, c'est-à-dire pour $\Delta x_r /c = 0$. Les accroissements d'indices *n* et *p* correspondants, $\Delta n$ et $\Delta p$, doivent donc obéir à

$$\Delta x_r /c = \Delta n. N_< - \Delta p. N_> = 0$$

ce qui impose

$$\Delta p / \Delta n = N_< / N_>, \text{ rapport fondamental noté } \chi.$$

Pour les mêmes $\Delta p$ et $\Delta n$ on a alors

$$\Delta x_b / c = \Delta n. N_> - \Delta p. N_<, \quad \Delta t_b = \Delta n. N_> + \Delta p. N_<,$$

et donc la vitesse réduite s'écrit

$$w = (\Delta x_b /c) / (\Delta t_b) = (\Delta n. N_> - \Delta p. N_<) / (\Delta n. N_> + \Delta p. N_<)$$
$$w = (1 - \chi^2) / (1 + \chi^2),$$

avec réciproquement

$$\chi = [(1 - w)/(1 + w)]^{1/2}$$

où l'on reconnaît le rapport Doppler, sous-jacent à toute cette analyse, dans son expression classique — et <u>irrationnelle</u> (cf. infra)—.

En éliminant *n* et *p* dans les expressions des coordonnées de $C_{n,p}$, on obtient

$$(t_r + x_r /c) / (t_b + x_b /c) = (t_b - x_b /c) / (t_r - x_r /c) = \chi \qquad \textbf{(L-E)}$$

écriture qui résume la transformation de coordonnées et fait simplement apparaître 1) l'"invariance du $ds^2$", puis, avec le caractère absolu-maximal de *c* 2) l'expression des coordonnées dans un repère d'après celles dans l'autre, 3) le caractère de groupe des transformations et 4) la composition des vitesses, le tout dans des termes — cette fois <u>rationnels</u> par $\chi$ — d'une remarquable simplicité.

Il peut paraître naïf d'insister sur le caractère "rationnel". Mais si le monde est quantique à la base, les notions les plus directes sont des entiers, puis des rapports d'entiers. Encore ces rapports ne sauraient-ils être simplifiés (rendus "irréductibles") arbitrairement, comme on verra. Il ne s'agit donc pas d'opposer une esthétique à une autre, un dogme à la Kronecker et un dogme à la Cantor, un finitisme à un infinitisme : il faut seulement prendre garde d'avancer en collant au caractère quantique, de sorte que les passages à de nouvelles variables et à certaines limites soient chaque fois contrôlés.

### 3. Quantum d'intervalle-action

Ce qui précède démontre que *la quantification des intervalles entre évènements implique les écritures de Lorentz-Einstein*. Il reste à prouver que <u>le</u> quantum naturel est la constante de Planck (<u>non pas</u> sa valeur réduite de $2\pi$), action et intervalle étant proportionnels : mais on aura un coefficient *positif*, et une nouvelle définition, naturelle, de l'action.

Pour cela, le point central de l'analyse tient dans la comparaison des figures 4 et 5. En figure 4 est reproduit le *pavé élémentaire* de la figure 0. Une ligne d'abscisse constante n'y est concrétisée que par les sommets correspondant aux émissions-absorptions de signaux : l'existence du "point matériel" de la droite physique n'est donc pas manifestée <u>entre</u> deux tels échanges, et cela n'est pas acceptable. Au contraire en figure 5 il y a tout un sous-programme de sommets intermédiaires (la suite éclairera que leur nombre relatif est élevé) au long de la ligne d'abscisse constante : ce sous-programme correspond en un sens au temps propre du grain matériel, en un autre sens à ce que ledit grain n'est plus considéré comme ponctuel.

A partir d'un pavé élémentaire ainsi autrement complexe, on peut cependant maintenir en vue de la traduction spatio-temporelle les flèches inclinées à $\pm \pi/4$, et concevoir un système physique comme un ensemble de signaux confinés. Ce contexte de traduction est marginal. Deux points au contraire sont fondamentaux.

L'essentiel, c'est le graphe, avec sa topologie. <u>De là les fréquences relatives</u> de flèches de chacun des types et, si on veut, leurs longueurs en représentations géométriques habituelles : car le *pavé* de base comporte alors, outre les *sommets* réels du sous-programme, des sommets <u>imaginaires</u> "à l'intersection des lignes portant les petites flèches" (en graphes, le décompte des valeurs relatives, intrinsèques, des nombres de flèches internes et externes aux sous-programmes suffit). Ces sommets imaginaires correspondent à un pavage plus fin, à partir duquel toutes les formules de la partie **2** précédente demeureront, le rapport $\chi$ et les paramètres spatio-temporels étant inchangés. Il faudrait disputer de l'intérêt de ces paramètres mais dès ce point il apparaît que, le rapport $\chi$ restant le même, <u>la répartition des coïncidences peut être considérablement moins dense et régulière que ce qui a été d'abord envisagé</u>, et les sommets intermédiaires réels considérablement moins nombreux que les imaginaires. En fait même, seulement <u>deux</u> coïncidences, l'une numérotée (*0, 0*) et l'autre (*n, p*), suffisent à ajuster une superposition de graphes-repères — ce qui affranchit de systèmes infinis et périodiques.

Outre ce premier point fondamental, on formule pour la suite l'hypothèse qu'un sous-programme définissant un système physique, un graphe-repère, doit comporter <u>un nombre égal de flèches de chacun des deux types</u> : en fait, cette hypothèse est directement liée à la définition ci-dessus de l'abscisse.

Ces deux essentiels posés, soit le graphe-repère propre d'un des systèmes (c'est-à-dire, dans la représentation <u>propre</u>, une allure comme fig. 0, mais où chaque pavé est <u>complété par un sous-programme dense</u> comme fig. 5) et soit une unité de temps assez grande pour que, le long d'une ligne *x/c = constante*, on trouve dans cette unité de très nombreux motifs du type de la figure 5. Soit alors $h$ la valeur attachée à <u>une</u> flèche, $\nu_0$ étant le nombre de flèches du sous-programme dans l'unité de temps choisie ; on pose

$$E_0 = \nu_0\, h$$

de sorte que, si $t_0$ est une durée mesurée dans cette unité, le nombre total de flèches du graphe-repère <u>*durant $t_0$*</u> est par définition

$$S = \nu_0\, h \cdot t_0 = E_0 \cdot t_0 \ .$$

Ce nombre, <u>lié au système</u> et à son graphe-repère, ne dépend pas du repère de décompte. On peut alors rappeler les formules (**L-E**) de Lorentz-Einstein ci-dessus (fin de titre **2**), à un jeu d'écriture près : en indices, on remplace $_r$ par $_0$ et on supprime simplement $_b$, en vue de différencier repère "propre" et "fixe" respectivement. D'où d'abord

$$t_0 = 1/2.\,(1/\chi + \chi)\, t - 1/2.\,(1/\chi - \chi)\, x/c$$
$$x_0/c = 1/2.\,(1/\chi + \chi)\, x/c - 1/2.\,(1/\chi - \chi)\, t \ .$$

Il suffit alors de multiplier par $\nu_0\, h$, et d'utiliser les abréviations

$$E = 1/2.\,(1/\chi + \chi)\, E_0 \qquad\qquad pc = 1/2.\,(1/\chi - \chi)\, E_0$$

pour obtenir d'abord les écritures en énergie-impulsion des formules de Lorentz-Einstein

$$E_0\, t_0 = E\, t - p\, x$$
$$E_0\, x_0/c = E\, x/c - pc\, t \ ,$$

puis successivement : l'interprétation de $p_0 = 0$ comme la demi-différence, par définition nulle, des nombres de flèches de chaque type pour un système physique ; les liens entre énergie et fréquence, entre longueurs d'onde de Compton et de Broglie puis impulsion, dans des conditions parallèles à celles des descriptions acquises. Il est alors nécessaire d'identifier $h$ à la constante de Planck, ce qui achève de fournir les éléments de la démonstration annoncée en titre et résumé.

### 4. Relecture

Dès 1905 donc (réf. [2]), Einstein ressentait la nécessité d'une justification directe de $E = h\nu$ : ici, on voit comment l'énergie n'est que l'indication d'une densité relative d'intervalles élémentaires lors de la comparaison de systèmes physiques, et il est clair que l'effet Doppler est plus physique que des discussions en termes de localisations, émissions et absorptions : *il n'y a pas de signaux, il n'y a que des relations, directes ou non*. Et l'abstraction, qui conduit à cette contradiction dans les termes qu'est la notion de « point matériel », est certes moins proche de l'expérience que celle d'*évènement* physique.

Dans la même voie, très vite ensuite, *l'intervalle* entre évènements apparaissait comme l'essentiel pour la structure intrinsèque, physique (cf. [3]). On peut même se demander alors comment, puisque cet intervalle était vu depuis longtemps comme proportionnel à l'action, et alors que la constante de Planck éclatait, si l'on peut dire, de toutes parts ([4]), tout ce qui précède n'a pas été trouvé depuis longtemps. En particulier, la vision d'Eddington sur un intervalle fondamental 0 ou 1 ([5]), surtout dans le contexte où il discute de la notion de métrique, est vraiment d'une pénétration remarquable ; et la méditation de de Broglie ([6]) indique nettement la nécessité, tout ensemble, de sortir du cadre devenu carcan spatio-temporel et de se référer à un point de départ « purement quantique ». C'est ce que synthétisait déjà Einstein ([7]), sauf peut-être que la qualification de « théorie purement <u>algébrique</u> » laisse de côté l'aspect <u>topologique</u>, indispensable pour le passage à la géométrisation au sens ordinaire. C'est un devoir de souligner de nouveau ici l'insondable valeur du testament scientifique d'Einstein ainsi relu : il renvoyait en vanité les efforts et aboutissements d'un demi-siècle de travaux, pourtant extraordinaires en histoire de la physique.

Plus remarquablement encore, Einstein voyait déjà plus loin même techniquement. Dans un texte médité en particulier par de Broglie, il écrivait (cf. *A. Einstein, philosopher and scientist* — P. A. Schilpp ed., *The Library of living Philosophers*, Open Court, 1949 — *Autobiographical notes* p. 63) : [trad.] « [...] la nature est ainsi faite qu'il est possible logiquement d'établir des lois assez fortement déterminées pour y faire intervenir des constantes complètement déterminées (et non, par conséquent, des constantes dont la valeur pourrait être changée sans détruire la théorie) ». On a vu ci-dessus comment les valeurs habituellement rapportées en $c$ et $h$ se trouvent ramenées à de simples unités de décomptes. Si toutefois les choses se développent comme il est actuellement envisageable, il y a fort à parier que la « troisième constante », dont la discussion peuple bon nombre d'articles depuis pratiquement un siècle, n'apparaîtra pas aussi simplement. L'idéal posé par Einstein accorde peut-être d'ailleurs une part trop belle à la façon dont la nature « <u>est faite</u> » : car c'est l'effort des physiciens de la <u>faire apparaître</u> comme régie par des lois dont la simplicité est chaque fois une surprise...

Pour en discuter plus précisément, il faudrait passer à la justification des trois dimensions de l'espace le plus apparent — cela du moins va vite à partir des n-simplexes —, montrer pourquoi la notion de champ est un lissage, transcrire la notion de force d'interaction, et surtout analyser plus avant la notion de présent : car l'insuffisance de ce qui précède ne tient pas tant à certaines non-transcriptions provisoires de beaucoup d'aspects expérimentaux, qu'au reliquat des <u>fautes</u> habituelles des paris spatio-temporels ; c'est cela qui rend la notion de présent presque aussi mal restituée que dans les descriptions habituelles — il suffirait d'inverser le sens de toutes les flèches pour obtenir un schéma de validité équivalente — et c'est cette insuffisance qu'il conviendra de réparer en vue d'autres élaborations. Mais que ceci déjà atteigne un commencement de public ; car d'autres résultats sont faciles à exhiber, et l'urgence est d'être accompagné par d'autres explorateurs dans cette redécouverte du cosmos. L'essentiel est donc de faire voir comment tout repart de simples doublets, évènements et intervalles quantiques, physiques, à l'encontre des invariants trompeurs et des entêtements spatio-temporels, avec leurs impossibles limites en dimensions très petites et autres renormalisations boiteuses. C'est cet essentiel qu'on a voulu avant tout communiquer.

# Figures

Légendes :

Fig. 0 : Schéma de relations ou signaux correspondant à une droite physique de points matériels équirépartis (repère propre)

Fig. 1 : Piquetage des coïncidences

Fig. 2 : Maille élémentaire "rouge"

Fig. 3 : Maille élémentaire "bleue"

Fig. 4 : Pavé élémentaire de sommets intermédiaires en géométrie du repère propre (cf. fig. 0) : la ligne d'abscisse constante (image du point matériel) <u>n'a pas</u> de structure interne

Fig. 5 : Pavé élémentaire de sommets intermédiaires en géométrie du repère propre : la "ligne" d'abscisse constante <u>correspond</u> cette fois à un <u>sous-programme</u>

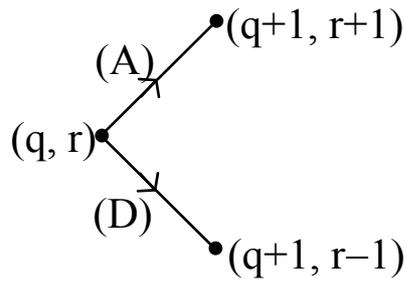

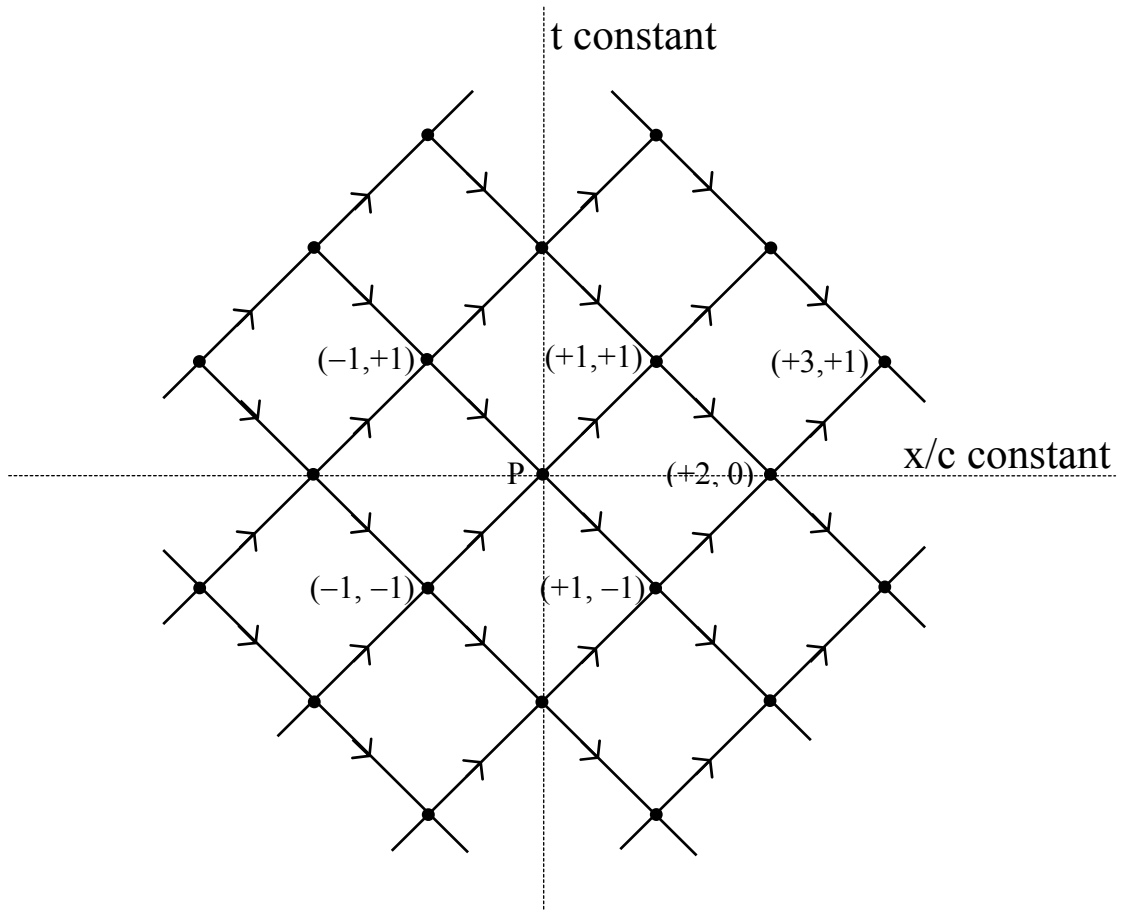

Fig. 0

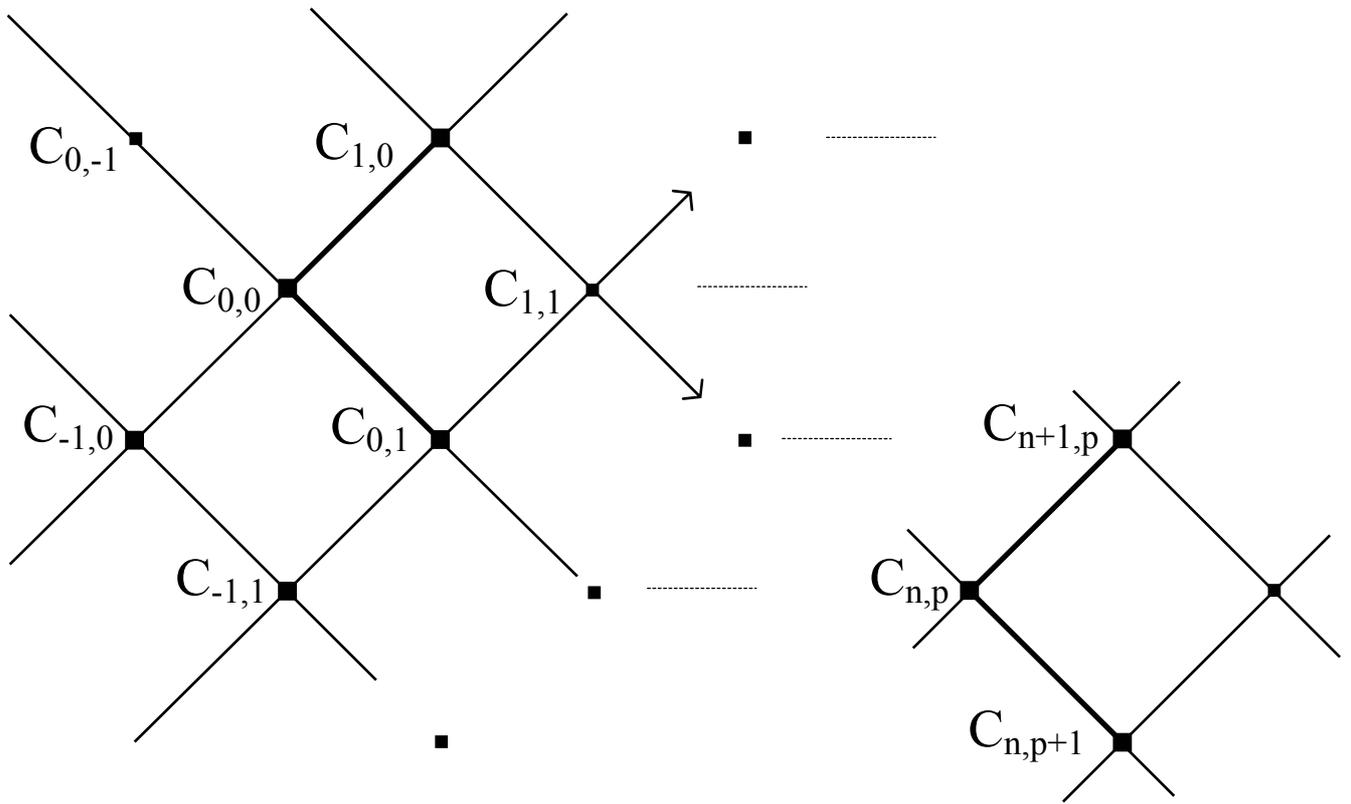

Fig. 1

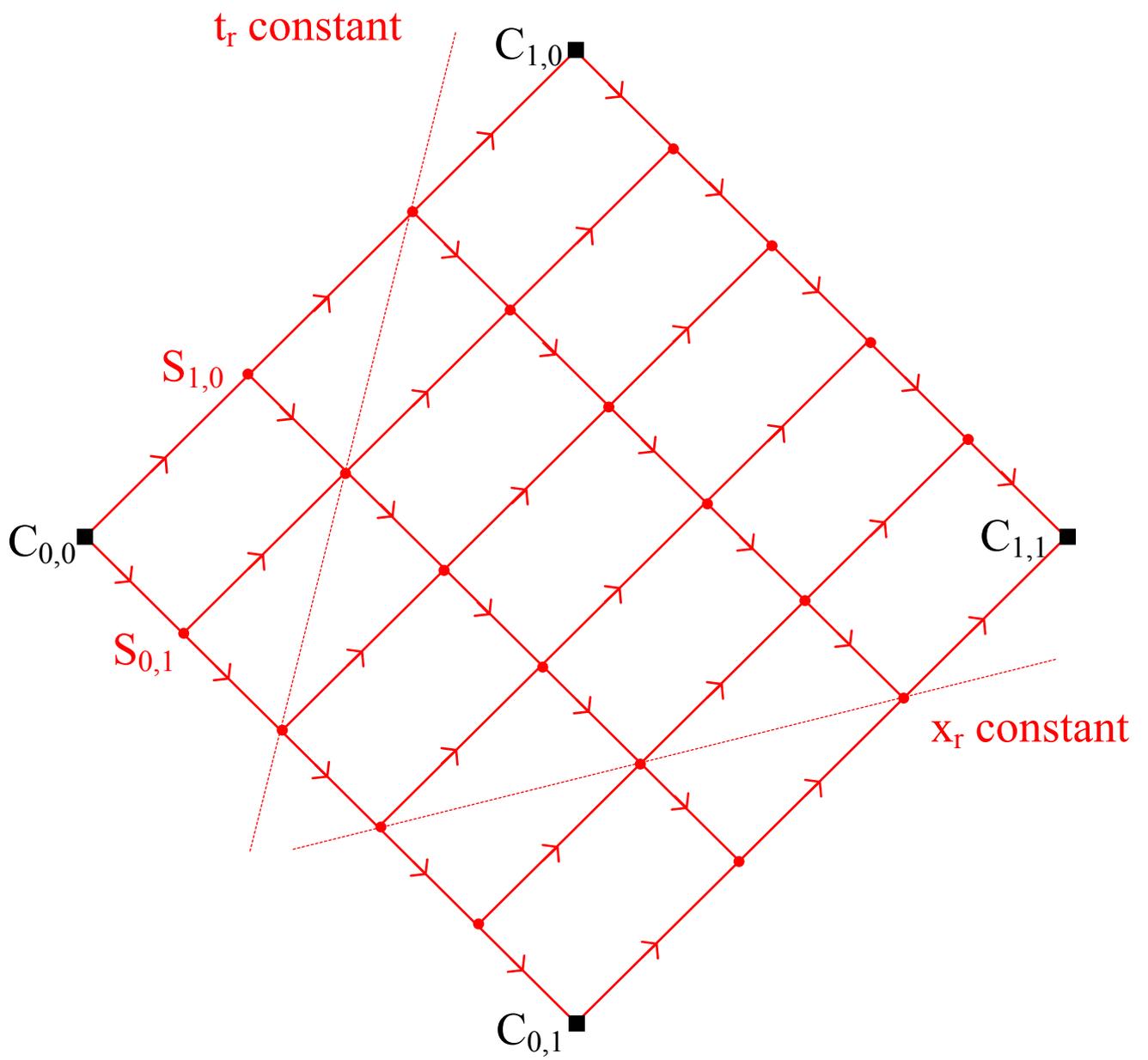

Fig. 2

($N_< = 3$, $N_> = 5$)

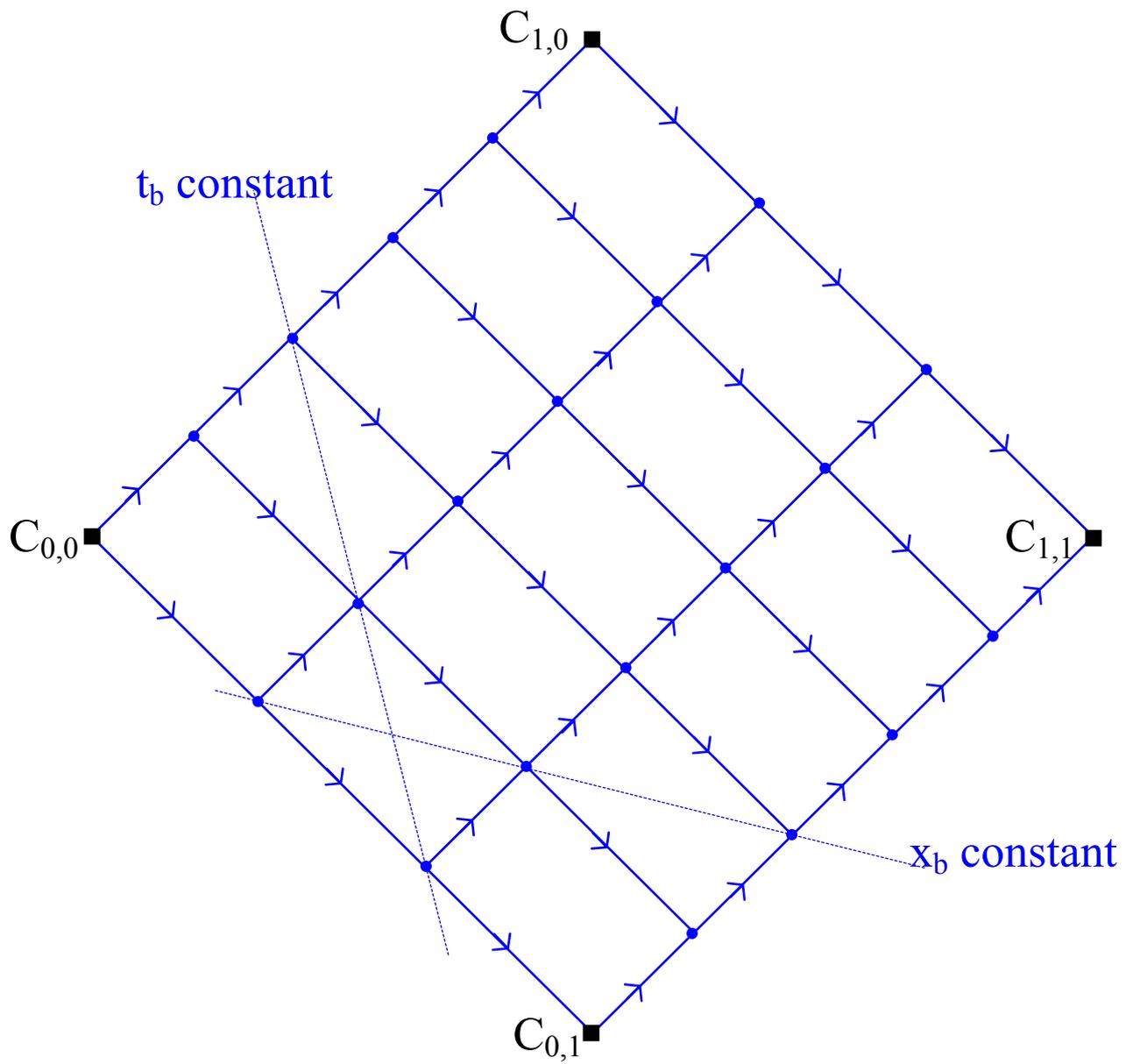

Fig. 3

(de nouveau $N_< = 3$,    $N_> = 5$  -  mais rôles permutés)

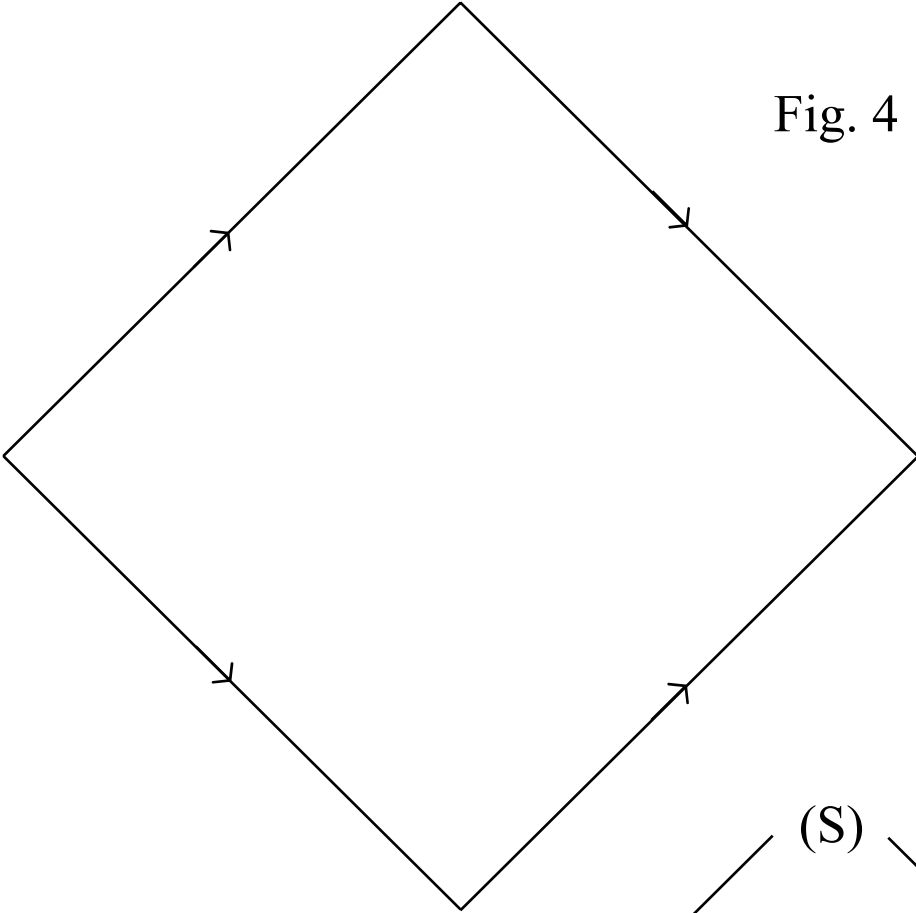

Fig. 4

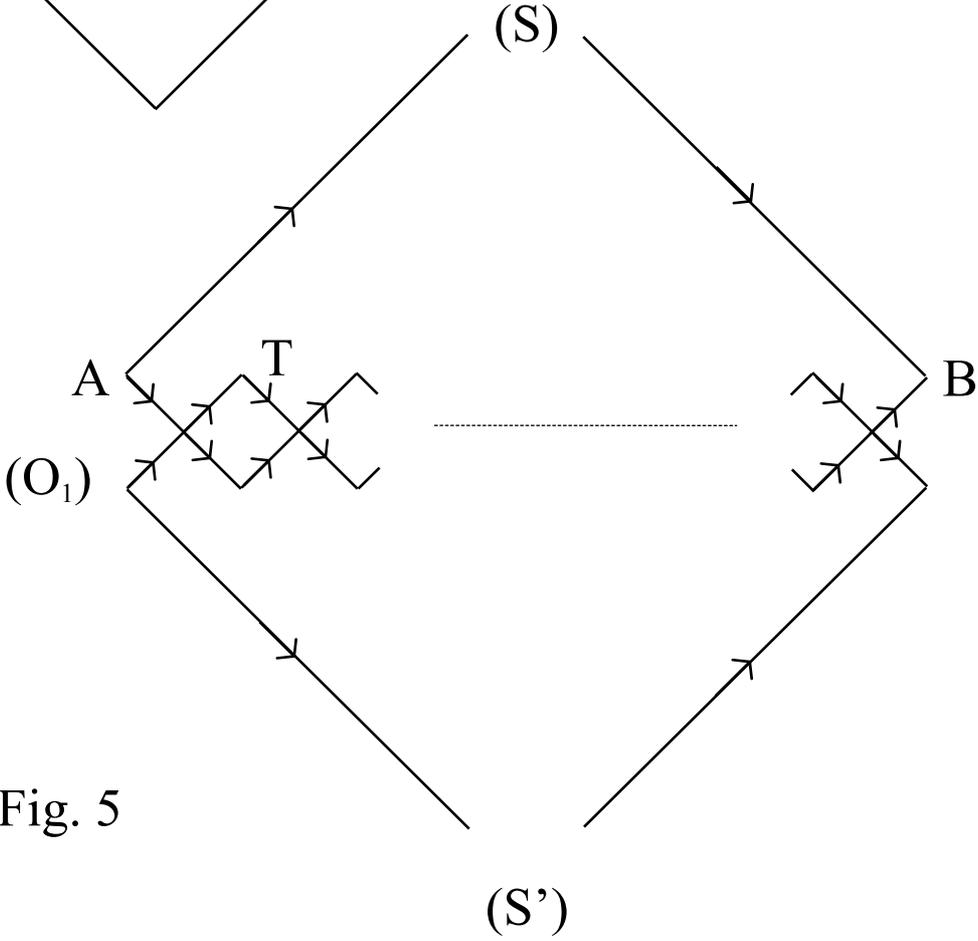

Fig. 5